\documentclass[journal]{IEEEtran}
\usepackage{cite}
\ifCLASSINFOpdf
\else
\fi
\usepackage{pgfplots}
\usepgfplotslibrary{fillbetween}
\usepackage{tikz}
\usetikzlibrary{fillbetween} 
\usetikzlibrary{patterns} 

\definecolor{myorange}{RGB}{255, 127, 80}
\definecolor{darkorange}{RGB}{200, 70, 30}
\definecolor{gridgray}{RGB}{230, 230, 230}
\definecolor{tealblue}{RGB}{0, 128, 128}
\definecolor{lightteal}{RGB}{102, 178, 178}

\usepackage{pgfplots}
\pgfplotsset{compat=1.18}
\usepackage{amsmath}
\usepackage{amssymb}
\usepackage{algorithm}
\usepackage{algpseudocode}
\usepackage{booktabs, multirow, array}
\usepackage{url}
\usepackage{graphicx}
\begin{document}

\title{Bidirectional Semantic Complementary Tool Retrieval for Remote Sensing Agents}

\author{
    Zeyuan~Wang$^{1}$, 
    Dongyang~Hou$^{1}$, 
    Cheng~Yang$^{1}$, 
    Xuezhi~Cui$^{1}$, 
    Linrui~Xu$^{1}$, 
    Bo~Yu$^{1}$, 
    Gaozhi~Zhou$^{2}$, \\
    Ziyu~Li$^{1}$, 
    Liangtian~Liu$^{1}$, 
    Kai~Ouyang$^{1}$, 
    Wang~Guo$^{1}$, 
    Lili~Zhu$^{3}$, 
    Chao~Tao$^{1}$
    
    \thanks{$^{1}$School of Geosciences and Info-Physics, Central South University, Changsha 410083, China.}
    \thanks{$^{2}$School of Mechanical and Electrical Engineering, Central South University, Changsha 410083, China.}
    \thanks{$^{3}$Hunan Key Laboratory of Land Resources Evaluation and Utilization, Hunan Provincial Institute of Land and Resources Planning, Changsha 410013, China.}
}

\markboth{}{...}
\maketitle

\begin{abstract}
Large language model (LLM)-based agents provide a novel paradigm for the automated processing of remote sensing (RS) data. Their success in complex RS tasks rely on extensive  specialized tool libraries. However, tool documentation often exceeds the context window limits of LLMs, making precise tool retrieval essential for agentic workflows. Existing tool retrieval methods face ``semantic asymmetry'' bottleneck: natural language queries typically express macro-level intentions lacking tool-specific semantics, while tool documentation provides fine-grained technical descriptions lacking operational context  for workflows. To bridge this semantic gap, this paper proposes a bidirectional semantic complementary tool retrieval method. First, on the query side, we introduce a planning-based query enhancement mechanism that leverages the reasoning capabilities of agents to decompose abstract intentions into logical sub-tasks, thereby actively supplementing the query with missing functional semantics. Second, on the tool side, addressing the strong coupling characteristics of RS tool chains, we construct a dynamic tool dependency graph with continual learning capabilities. By employing a neighborhood information aggregation mechanism, contextual information from precursor tools is explicitly injected into the current node representation, enriching tool descriptions with contextual semantics. Experimental results on the RS dataset GeoPlan-bench and the general-purpose dataset API- Bank demonstrate that the proposed method not only significantly improves tool retrieval accuracy for complex RS tasks but also exhibits robust extensibility for transfer to general-domain tasks. The source code and dataset are available at https://github.com/geox-lab/BSCTR.
\end{abstract}


\IEEEpeerreviewmaketitle
\begin{figure*}[!t]
    \centering
    \includegraphics[width=1.0\textwidth]{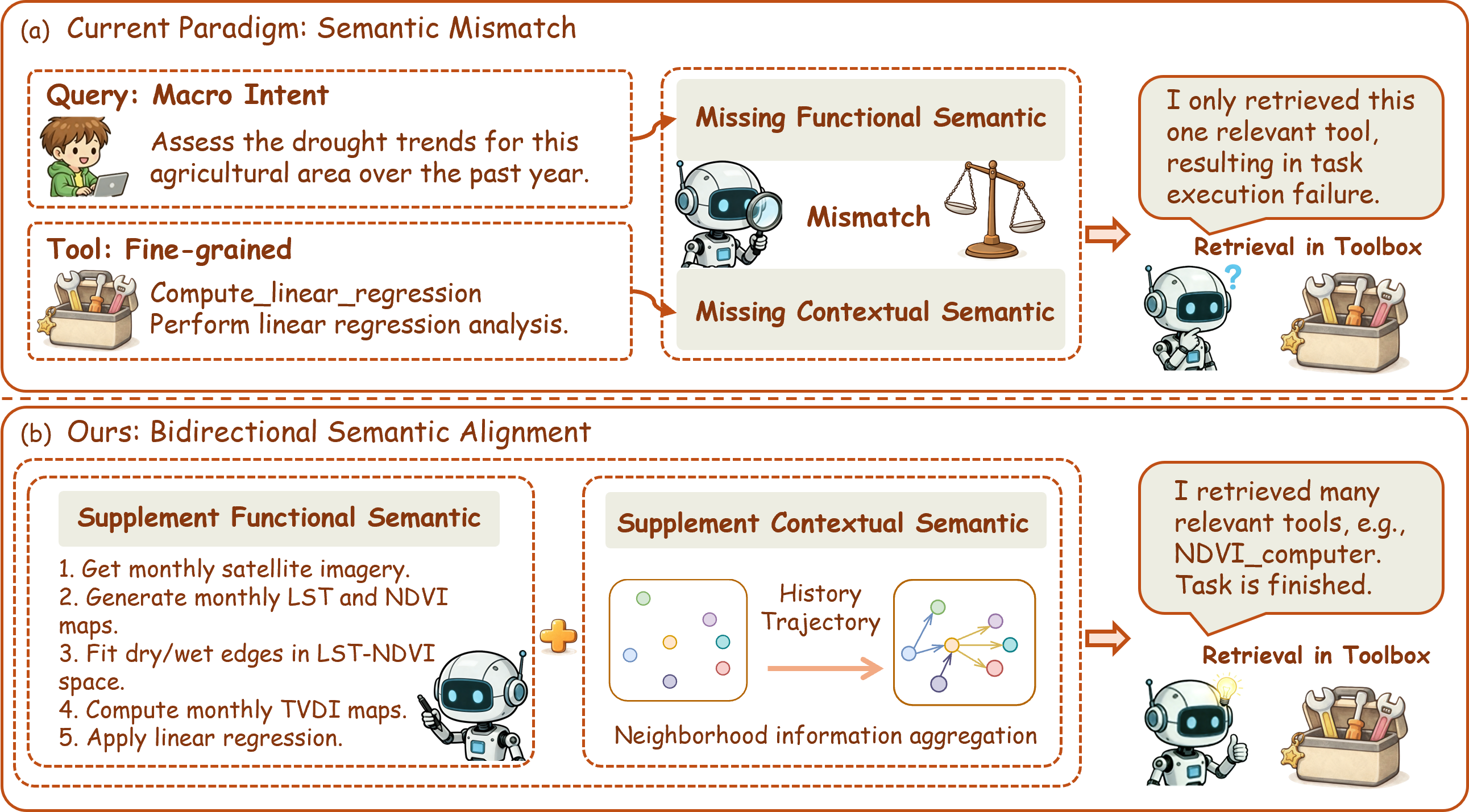} 
    
    \caption{Comparison of tool retrieval paradigms for the remote sensing agent. 
    (a) \textbf{Current Paradigm}: Semantic mismatch occurs due to missing functional and contextual semantics, leading to task execution failure. 
    (b) \textbf{Ours}: Bidirectional semantic alignment effectively bridges the gap by supplementing both semantics, ensuring successful retrieval of all necessary tools.}
    \label{fig:motivation}
\end{figure*}

\section{Introduction}
\IEEEPARstart{W}{ith} the rapid advancement of satellite technology, Earth observation capabilities have been markedly improved, leading to a marked increase in image acquisition efficiency. However, the efficiency of image interpretation is far behind, exacerbating the contradiction between ``massive data'' and``scant information''. Faced with multi-modal remote sensing (RS) observation data with varying spatio-temporal resolutions, diverse application scenarios impose distinct requirements on interpretation tasks. Consequently, systems must be capable of invoking specific combinations of models and algorithms to precisely address these varying needs. Nevertheless, in current practice, automated workflows relying on expert-defined templates struggle to flexibly adjust tool selection and composition according to dynamic task requirements, which has become a major bottleneck restricting the automation of RS interpretation. 

To overcome this limitation, Large Language Model (LLM)-driven agents have emerged as a promising research direction. Benefiting from breakthroughs in intent understanding, autonomous decision-making, and tool invocation capabilities\cite{react,plansolve}, LLM-based agents can flexibly handle complex scenarios that traditional workflows find difficult to adapt to through logical reasoning. Accordingly, researchers in the RS field have begun to introduce the agent paradigm. For instance, the RS-Agent team proposed RS-Agent\cite{rsagent}, which aims to achieve autonomous scheduling of specialized models through natural language interaction. For more complex long-chain tasks, the Earth-Agent team\cite{geoplan} proposed a hierarchical multi-agent architecture to achieve task planning and execution following domain-specific processing workflows. These studies demonstrate that agents, with their flexible reasoning and planning capabilities, are providing a novel processing paradigm for RS interpretation tasks.

However, remote sensing agents still face severe challenges. RS image analysis tasks typically involve highly specialized tool-chains, encompassing a series of intricate steps from data preprocessing and radiometric calibration to geometric correction and feature extraction. To tackle complex image analysis tasks, agents must be equipped with extensive open-source tool libraries—such as QGIS, Orfeo ToolBox (OTB), and GDAL—which comprise hundreds of specialized functions. In early practices, researchers tended to adopt a ``full-tool registration'' strategy, where functional descriptions of all available tools were directly injected into the agent's system prompts\cite{toollearning}. However, constrained by the finite context window capacity of LLMs, excessive tool descriptions often lead to a significant degradation in the agent's task execution performance \cite{Langchain}. Consequently, when dealing with extensive tool-sets, the tool retrieval paradigm has emerged as the mainstream approach. This paradigm dynamically recalls the most relevant tool subset by calculating the semantic similarity between user queries and tool descriptions, thereby facilitating precise tool supply within a limited context \cite{Gorilla,toolllm}.

Although retrieval methods mitigate the aforementioned capacity constraints, their accuracy is significantly hindered by the ``\textit{semantic misalignment}'' between user queries and tool descriptions. Essentially, tool retrieval is a task of alignment within the semantic space. Drawing upon the foundational theories of semiotician Charles W. Morris \cite{morris}, we redefine the semantics of tool retrieval in the context of agent-based tool invocation into two dimensions: ``\textit{functional semantics}'' and ``\textit{contextual semantics}.'' Specifically, functional semantics focus on a tool's technical implementation and execution objects, whereas contextual semantics emphasize its role and usage within a specific task workflow. Existing retrieval paradigms face a distinct ``\textit{semantic misalignment}'': user queries are typically coarse-grained macro-intentions that lack the functional semantic descriptions of specific technical means; conversely, tool documentation provides fine-grained technical specifications that define inputs and outputs but lack the contextual semantics of real-world application scenarios(as conceptually illustrated in Fig. \ref{fig:motivation}(a)). Current literature has only addressed this issue from a single perspective. For instance,\cite{toolreagt,improving} focus on the query side by rewriting user queries to supplement functional semantics, while \cite{toolnet,graph-toolfusion} construct graph connections to enrich contextual information on the tool side. However, these efforts fail to fully bridge the fundamental semantic gap that exists between query descriptions and tool specifications.

Within the field of RS, we observe distinct ``strong coupling'' characteristics among specialized tools. Specific steps in RS tasks typically adhere to established operational paradigms; for instance, atmospheric correction is usually followed by geometric correction. Consequently, the outputs of precursor tools often impose hard constraints on the selection of subsequent tools. We contend that these domain-specific tool-chain dependencies constitute the contextual semantics of a tool—that is, the profound meaning of a tool is inherently embedded within the context of its entire operational workflow. Building upon this hypothesis, this paper proposes a \textbf{Bidirectional Semantic Complementary Tool Retrieval} (BSCTR) method, which effectively aligns the missing semantics from both sides (see Fig. \ref{fig:motivation}(b)). On the query side, leveraging the robust reasoning and planning capabilities of contemporary agents \cite{plansolve}, we introduce a planning-based query enhancement mechanism. This mechanism drives the agent to logically decompose and reason through coarse-grained user intentions, autonomously excavating and supplementing missing functional semantics within the query. On the tool side, we construct a dynamic tool dependency graph with continual learning capabilities. Initially, topological connections between tools are initialized based on historical trajectories. Then, inspired by the information aggregation mechanism of Simplified Graph Convolution \cite{sgc}, we perform linear aggregation of neighborhood features in the vector space. This explicitly injects contextual information from precursor tools into the current tool node representations, thereby enriching the contextual semantics of tool descriptions. To ensure the graph's generalization capability, the system autonomously establishes new connections between tools invoked sequentially in successful task executions. This bi-directional enhancement effectively bridges the semantic misalignment between user queries and tool descriptions.

The primary contributions of this work are summarized as follows:
\begin{itemize}
    \item \textbf{Bidirectional Complementary Framework:} We propose the \textbf{Bidirectional Semantic Complementary Tool Retrieval} framework, which establishes a novel paradigm for bridging the ``semantic asymmetry'' in RS agent tasks. By synergizing planning-based query enhancement and tool-side contextual injection, the framework effectively aligns macro-level user intentions with fine-grained technical specifications.
    
    \item \textbf{Graph-based Contextual Modeling:} We design a dynamic tool dependency graph mechanism that explicitly encodes pragmatic contexts into node representations. Through a neighborhood information aggregation mechanism in the vector space, the framework injects precursor contextual semantics into tool descriptions, capturing the strong coupling characteristics and sequential logic inherent in professional RS workflows.
    
    \item \textbf{Empirical Validation:} We validate the superiority of the proposed method on the specialized RS dataset \textit{GeoPlan-bench} \cite{geoplan}. Moreover, evaluations on the general-purpose \textit{API-Bank} \cite{api-bank} demonstrate the robust domain scalability and generalization capabilities of our mechanism.
\end{itemize}


\section{Related work}
\begin{figure*}[!t]
\makebox[\textwidth][c]{
\includegraphics[width=1.0\textwidth]{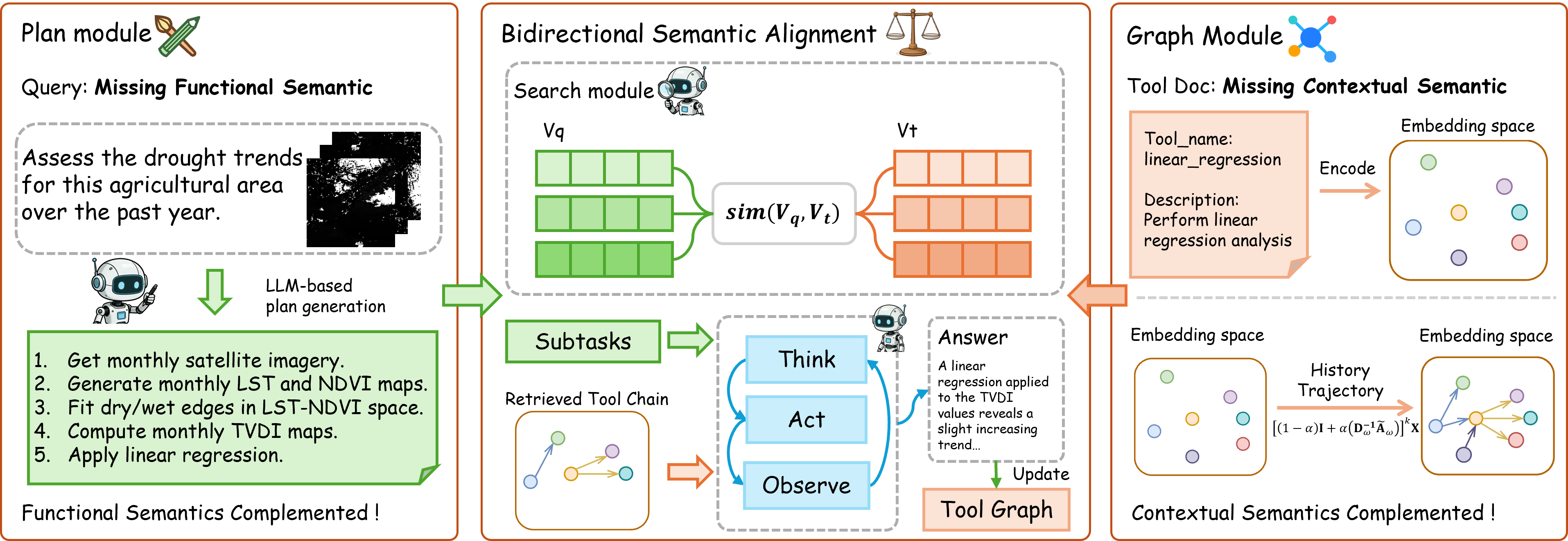}
}
\caption{Overall framework of our proposed bidirectional semantic alignment approach. The \textbf{Plan module} (left) complements functional semantics by generating LLM-based subtasks from user queries. The \textbf{Graph module} (right) complements contextual semantics by encoding tool documents and aggregating historical trajectories. Finally, the \textbf{Bidirectional Semantic Alignment module} (center) performs precise tool retrieval by matching representations ($V_q$ and $V_t$) and executes the tasks via a reasoning loop to output the final answer.}
\label{fig:framework}
\end{figure*}
This section systematically reviews the research progress of large language models (LLMs) within the remote sensing (RS) domain and the field of tool selection from three distinct dimensions. Subsection \ref{subsec:rs_agents} traces the evolutionary trajectory of RS agents, elucidating the paradigm shift from early multi-modal perception to advanced intelligent systems equipped with logical planning and autonomous tool selection capabilities. Subsection \ref{subsec:query_rewriting} investigates tool retrieval strategies based on query rewriting, analyzing how researchers supplement missing functional semantics on the query side by reformulating user intentions. Subsection \ref{subsec:tool_graph} summarizes tool retrieval methods leveraging tool graphs, exploring the current landscape of enriching tool contextual semantics through the modeling of topological dependencies. 

\subsection{Remote sensing agents based on large language models}
\label{subsec:rs_agents}
In recent years, the integration of LLMs and multi-modal technologies has reshaped the processing paradigms for RS imagery. Early research primarily focused on constructing RS multi-modal foundation models to enhance the perception and understanding of domain-specific knowledge. For example, \textbf{GeoChat} \cite{geochat} empowered RS models with visual grounding capabilities through instruction tuning, achieving the alignment of object descriptions with spatial coordinates. \textbf{EarthGPT} \cite{earthgpt} addressed the domain gaps between optical, Synthetic Aperture Radar (SAR), and infrared imagery by integrating multi-source sensor data. Building on these advancements, \textbf{AllSpark} \cite{allspark} proposed the concept of ``language as a reference frame'', mapping over ten types of heterogeneous spatio-temporal modalities---including RS images, point clouds, and trajectories---into a unified linguistic feature space to achieve cross-modal general spatio-temporal intelligence.

These foundation models established a solid repository of domain knowledge for subsequent logical reasoning in intelligent agents. However, static perception models struggle to handle complex interactive RS analysis tasks, leading to a research shift from ``visual perception'' toward \textbf{autonomous agents}. These agents utilize the LLM as a core decision-making brain to solve practical engineering problems through task planning and tool invocation. In terms of interaction and execution, \textbf{RSAgent} \cite{rsagent} demonstrated the autonomous scheduling of specialized models through natural language interaction to satisfy practical RS requirements. Furthermore \textbf{Earth-Agent}\cite{earthagentbench} designed specialized toolsets for spectral, RGB, and product-oriented tasks, aiming to empower agents with the capability to handle a diverse range of applications within the remote sensing domain. For complex task planning, \textbf{EarthAgent} \cite{geoplan} introduced a hierarchical multi-agent architecture and developed the GeoPlan-bench benchmark, proving the potential of agent systems in executing multi-step planning that follows domain-specific processing workflows. Additionally, \textbf{GeoGPT} \cite{geogpt} leveraged a ``Think-Act-Observe'' closed-loop mechanism to transform natural language instructions into professional Geographic Information System (GIS) workflows. 

Despite these advancements, a critical bottleneck emerges when agents are required to interface with massive open-source tool libraries such as GDAL, QGIS, and OTB. Due to the inherent constraints of LLM context window capacities, the conventional ``full-tool registration'' strategy---which involves injecting exhaustive functional descriptions of all tools into the system prompt---becomes unfeasible and leads to significant performance degradation. Consequently, the paradigm for agent tool management has inevitably transitioned from exhaustive registration toward \textbf{retrieval-based registration}, where the most relevant tool subset is dynamically recalled to maintain precise tool supply within a limited context.

\subsection{Query rewriting-based tool retrieval}
\label{subsec:query_rewriting}
Leveraging the formidable language understanding and task planning capabilities of LLMs \cite{cot, llm-plan, tot}, researchers have increasingly utilized the innate reasoning of LLMs to rewrite coarse-grained user queries into fine-grained representations, which are subsequently matched with tool documentation via standard retrieval paradigms, such as sparse lexical matching (e.g., BM25) or dense semantic similarity.

Specifically, \textbf{ToolReAGt} and \textbf{Kachuee \textit{et al.}} \cite{toolreagt, improving} employ query rewriting to supplement functional semantics within user queries. While \textbf{Kachuee \textit{et al.}} \cite{improving} introduces three distinct strategies to enhance the rewriting capabilities of LLMs, \textbf{ToolReAGt} \cite{toolreagt} integrates the \textbf{ReAct} \cite{react} execution paradigm to observe tool processing results and determine subsequent actions post-retrieval. \textbf{Reinvoke} \cite{reinvoke} adopts a bi-directional rewriting approach: it generates fine-grained rewrites of user queries while simultaneously ``imagining'' potential problems that specific tools might solve, performing semantic matching between these two augmented representations. Similarly, \textbf{ToolDreamer} \cite{tooldreamer} constructs a contrastive dataset comprising LLM-imagined tool descriptions versus real-world tool specifications to train a specialized retriever, aiming to align the discrepancies between LLM hallucinations and actual tool functionalities. Although these works effectively supplement the missing functional semantics on the query side, they overlook the inherent lack of \textbf{contextual semantics} in the tool descriptions themselves.
\subsection{Tool retrieval based on tool graph}
\label{subsec:tool_graph}
Given the task-driven inherent connections between tools, recent research has increasingly focused on modeling tool structures to enhance retrieval performance. Based on how the graph structure is utilized, existing work can be categorized into two paradigms: \textbf{explicit topological traversal} and \textbf{implicit feature aggregation}.

\textbf{Explicit topological traversal} aims to leverage graph connections to expand the retrieval scope. \textbf{ToolNet} \cite{toolnet} constructs a tool graph with dynamically updated weights based on historical trajectories, aiming to recall tools in the form of tool-chains while providing a reference for agent tool selection through edge weights. \textbf{Graph RAG-Tool Fusion} \cite{graph-toolfusion} defines inter-tool dependencies and employs a strategy that retrieves seed nodes and then utilizes a Depth-First Search (DFS) algorithm to traverse the graph for candidate tool selection. \textbf{AutoTool} \cite{autotool} adopts a dynamic graph construction approach; after selecting the current tool, it employs a dual-layer filtering mechanism to bypass LLM reasoning for subsequent executions, thereby reducing token consumption. Furthermore, \textbf{SciToolAgent} \cite{scitoolagent} utilizes manually constructed scientific tool knowledge graphs to guide LLMs in planning and executing multi-step tool-chains in domains such as biology, chemistry, and materials science. Fundamentally, these methods treat the graph as a post-retrieval search path, making their effectiveness highly dependent on the accuracy of the initial ``seed node'' retrieval. In scenarios with ambiguous user intent, failing to hit the correct seed node initially causes the subsequent graph traversal to fall into incorrect search branches.

\textbf{Implicit feature aggregation}, on the other hand, attempts to integrate graph structural information directly into tool representations. For instance, \textbf{TGR} \cite{tgr} trains a tool connection generator to predict whether dependencies exist between tool pairs, subsequently employing Graph Convolutional Networks (GCNs) \cite{gcn} to aggregate features and capture inter-tool correlations within the vector space. However, the quality of the resulting tool graph in such approaches is strictly constrained by the predictive performance of the connection generator. The method proposed in this paper also falls under the paradigm of implicit feature aggregation, yet it diverges from \textbf{TGR} in several key aspects. First, instead of relying on predictive models, we construct the tool dependency graph based on \textbf{historical trajectories}, thereby directly capturing authentic tool invocation logic and enabling real-time graph updates. Second, we uniquely supplement the \textit{functional semantics} of tools on the user query side. By doing so, our approach avoids the heavy reliance of explicit traversal on ``seed nodes'' and achieves dynamic, semantically enriched tool context aggregation.

\section{Methodology}
\label{sec:methodology}
This section delineates the proposed \textbf{Bidirectional Semantic Complementary Tool Retrieval (BSCTR)} framework in detail. The framework is designed to bridge the semantic misalignment between user natural language queries and specialized tool descriptions by performing \textit{functional semantic completion} on the query side and \textit{contextual semantic injection} on the tool side. As illustrated in Fig. \ref{fig:framework}, the system architecture is logically organized into three primary modules:

\begin{enumerate}
    \item \textbf{Plan Module (Query-side Functional Enhancement):} This module leverages the reasoning capabilities of LLMs to decompose abstract user intentions into logical sub-tasks, thereby actively excavating and supplementing the missing functional semantics to generate enhanced query representations ($V_q$).
    
    \item \textbf{Graph Module (Tool-side Contextual Injection):} To capture the strong coupling features of remote sensing (RS) tool-chains, this module encodes tool documents and aggregates historical trajectories. By employing a graph-based topological neighborhood aggregation mechanism, it explicitly injects contextual information from precursor tools into current node representations ($V_t$).
    
    \item \textbf{Bidirectional Semantic Alignment Module (Retrieval, Execution, and Evolution):} Serving as the core operational hub, this module performs precise tool retrieval by computing the similarity between the enhanced representations ($V_q$ and $V_t$). Furthermore, it executes the retrieved sub-tasks via a continuous reasoning loop to output the final answer. Crucially, based on the environmental feedback from this reasoning loop, the module triggers a dynamic tool graph evolution mechanism, autonomously establishing and updating tool dependencies from successful task trajectories to ensure the tool graph's continual adaptability.
\end{enumerate}

\subsection{Plan Module (Query-side Functional Enhancement)}
\label{subsec:query_enhancement}

To address the deficiency of tool functional semantics in user queries, this section introduces a planning-based enhancement strategy leveraging the reasoning power of LLMs. 
The core of this strategy is to decompose abstract user goals into a specific sequence of sub-tasks characterized by explicit functional semantics. 

Given a coarse-grained user query $Q_{coarse}$, we utilize the task planning capabilities of LLMs (e.g., the Plan-and-Solve prompting mechanism \cite{plansolve}) to decompose the high-level intention into a set of logical sub-tasks $S$, defined as:
\begin{equation}
S = \{s_1, s_2, \dots, s_n\}
\label{eq:subtasks}
\end{equation}
For each specific sub-task $s_i \in S$, we fuse it with the original macro-intention to construct a sub-task-specific enhanced query $q_{enhance}^{(i)}$:
\begin{equation}
q_{enhance}^{(i)} = Q_{coarse} \oplus s_i
\label{eq:q_enhance}
\end{equation}
where $\oplus$ denotes the textual concatenation operation. 

Through this mechanism, each sub-task---which originally lacked broad context---is injected with rich functional semantics from the macro-query, forming the final query representations $V_q$. This significantly improves the alignment and matching accuracy between the query and tool documentation within the high-dimensional vector space.

\subsection{Graph Module (Tool-side Contextual Injection)}
\label{subsec:tool_graph_modeling}

To address the lack of contextual semantics in tool descriptions, this section introduces a method to construct a tool dependency graph and perform multi-order feature aggregation. 
This approach explicitly injects the ``strong coupling'' relationships between tools into their node representations, thereby enriching the tool descriptions with pragmatic contextual information.

We first define the tool library as a directed graph $G = (N, E)$, where $N$ denotes the set of original tools in the library and $E$ represents the sequential invocation dependencies. Specifically, a directed edge $(i, j) \in E$ indicates that tool $i$ is a precursor invoked immediately before tool $j$. Accordingly, we define the adjacency matrix $\mathbf{A}$, where $\mathbf{A}_{ij} = 1$ if $(i, j) \in E$, and $0$ otherwise.

To flexibly capture tool contexts from different propagation directions, we introduce a direction indicator variable $\omega \in \{p, s, sym\}$ and formulate the corresponding augmented adjacency matrix $\mathbf{\tilde{A}}_\omega$ as follows:
\begin{equation}
\mathbf{\tilde{A}}_\omega = 
\begin{cases} 
\mathbf{A}^T + \mathbf{I}, & \text{if } \omega = p \\ 
\mathbf{A} + \mathbf{I}, & \text{if } \omega = s \\ 
\mathbf{A} + \mathbf{A}^T + \mathbf{I}, & \text{if } \omega = sym 
\end{cases}
\end{equation}
where $\mathbf{I}$ is the identity matrix representing self-connections. Specifically: 1) \textit{Predecessor Mode} ($\omega = p$) aggregates historical information from precursor tools (i.e., incoming edges); 2) \textit{Successor Mode} ($\omega = s$) aggregates future information from subsequent tools (i.e., outgoing edges); and 3) \textit{Symmetric Mode} ($\omega = sym$) aggregates undirected contextual information from both directions.

For normalized feature diffusion, we define the degree matrix $\mathbf{D}_\omega$ as a diagonal matrix where each diagonal element is the row sum of $\mathbf{\tilde{A}}_\omega$:
\begin{equation}
(\mathbf{D}_\omega)_{ii} = \sum_{j=1}^{|N|} (\mathbf{\tilde{A}}_\omega)_{ij}
\end{equation}
Inspired by the Simplified Graph Convolutional Network (SGC) \cite{sgc} and the Personalized PageRank (PPR) propagation algorithm \cite{ppr}, we utilize a multi-order linear aggregation operator based on the matrix pair $\{\mathbf{\tilde{A}}_\omega, \mathbf{D}_\omega\}$ to map the initial tool features $\mathbf{X}$ into a pragmatic enhancement space. The $k$-th order pragmatically enhanced feature matrix $\mathbf{H}^{(k)}$ is computed as:
\begin{equation}
\mathbf{H}^{(k)} = \left[ (1-\alpha)\mathbf{I} + \alpha \left( \mathbf{D}_\omega^{-1} \mathbf{\tilde{A}}_\omega \right) \right]^k \mathbf{X}
\end{equation}
where $\alpha \in [0, 1]$ is a smoothing factor used to balance the tool's original \textit{functional semantics} and its \textit{contextual semantics} from topological neighborhoods. 
By adjusting the order $k$, tool nodes can selectively assimilate pragmatic features from task chains of varying lengths, thereby enriching the contextual semantics of tool descriptions. The row vectors of the resulting pragmatically enhanced feature matrix $\mathbf{H}^{(k)}$ constitute the injected contextual representations $V_t$ for all tools.

\subsection{Bidirectional Semantic Alignment Module (Retrieval, Execution, and Evolution)}
\label{subsec:alignment_module}

To operationalize the proposed bidirectional semantic complementarity mechanism, we formalize the system's logic into a closed-loop process within the Bidirectional Semantic Alignment module. We define the overall agent system interacting through this central hub as a quintuple:
\begin{equation}
\Omega = \langle \Pi, \mathcal{G}, \mathcal{S}, \mathcal{A}, \mathcal{U} \rangle
\end{equation}
where $\Pi$ denotes the query enhancement module, $\mathcal{G}$ is the tool graph module, $\mathcal{S}$ represents the retrieval module, $\mathcal{A}$ is the task execution module, and $\mathcal{U}$ indicates the graph update module. The operational workflow of this module is logically partitioned into three progressive phases, as summarized in Algorithm \ref{alg:agent_process}.

\subsubsection{Initialization and Retrieval}
Prior to deployment, the agent initializes the tool graph using offline historical trajectory data. By traversing tool-chains and analyzing invocation dependencies, an initial topological structure $\mathcal{G}_{init}$ is established, and the pragmatically enhanced representations $\mathbf{H}^{(k)}$ for each tool node are computed. This stage provides the essential contextual semantics required for retrieval during the cold-start phase.

During the online inference phase, upon receiving a user query $Q$, the agent invokes the planning mechanism to decompose the task into a logical sub-task sequence $S = \{s_1, s_2, \dots, s_n\}$. For each sub-task $s_i$, the agent utilizes module $\Pi$ for query enhancement to generate $q_{enhance}^{(i)}$. Based on the enhanced sub-query $q_{enhance}^{(i)}$ and the tool graph $G$, the retrieval module $\mathcal{S}$ calculates the cosine similarity between the sub-query and the tool node representations. The top-$k$ relevant tools for each $s_i$ are recalled to form a candidate subset $T_i$. Tools from all subsets are then globally re-ranked using the Reciprocal Rank Fusion (RRF) algorithm \cite{rrf} and stored in a consolidated tool candidate pool $Pool_T$.

\subsubsection{Execution via Reasoning Loop}
Subsequently, the task execution module $\mathcal{A}$ adopts the ReAct architecture \cite{react} to execute the sub-tasks sequentially. During each step, the agent selectively invokes a tool from the limited context of $Pool_T$ and observes its execution outcome from the environment. Successful invocation steps and their logical sequence are meticulously recorded in the current tool invocation trajectory $P = \{t_1, t_2, \dots, t_m\}$. This reasoning loop continues until the final task objective is achieved.

\subsubsection{Dynamic Tool Graph Evolution}
To facilitate the continual learning of RS-domain-specific tool coupling logic, the module features a dynamic evolution mechanism based on successful execution feedback. Upon task completion, the agent evaluates the execution sequence $P$ and triggers the update module $\mathcal{U}$ through the following steps:
\begin{itemize}
    \item \textbf{Trajectory Collection and Verification:} If the task successfully passes verification through environmental feedback, the recorded trajectory $P$ is marked as a ``candidate for structural learning.''
    
    \item \textbf{Topological Connection Update:} For any adjacent tool pair $(t_i, t_{i+1})$ within the successful trajectory $P$, the system evaluates its existence in the current tool graph $\mathcal{G}$. If the pair represents a newly emerged combination, a directed edge $e_{i,i+1}$ is dynamically added to the graph.
    
    \item \textbf{Information Aggregation Update:} Following the topological update of $\mathcal{G}$, the aggregation operator defined in Section \ref{subsec:tool_graph_modeling} is re-invoked on the updated graph structure. Consequently, the pragmatically enhanced feature vectors $\mathbf{H}^{(k)}$ undergo evolution and migration as task experience accumulates.
\end{itemize}

This dynamic evolution mechanism empowers the tool library with \textbf{continual learning} capabilities, enabling the system to robustly handle complex remote sensing tasks in open-set environments.
\begin{algorithm}[t]
\caption{Agent Reasoning and Knowledge Evolution}
\label{alg:agent_process}

\makeatletter
\newcommand{\PState}[1]{\State \parbox[t]{\dimexpr\linewidth-\ALG@thistlm}{\strut #1\strut}}
\makeatother

\begin{algorithmic}[1]
\Require \parbox[t]{\dimexpr\linewidth-4em}{Historical trajectories $\mathcal{T}_{off}$, Initial tool features $\mathbf{X}$, User queries $\mathcal{Q}$}
\Ensure \parbox[t]{\dimexpr\linewidth-4em}{Updated tool graph $G$, Pragmatically enhanced features $\mathbf{H}^{(k)}$}

\Statex \textbf{// Phase 1: Initialization}
\PState{Construct initial graph $G$ using $\mathcal{T}_{off}$} 
\PState{Compute pragmatic features $\mathbf{H}^{(k)}$ using $\mathbf{X}$ and $G$ via Eq. (5)}

\Statex \textbf{// Phase 2 \& 3: Inference Loop and Dynamic Evolution}
\For{each query $Q_{coarse} \in \mathcal{Q}$}
    \PState{Decompose $Q_{coarse}$ into sub-tasks $S = \{s_1, \dots, s_n\}$ via Planner} 
    \PState{Initialize Tool Candidate Pool $Pool_T \leftarrow \emptyset$}
    \For{each sub-task $s_i \in S$}
        \PState{Construct enhanced query $q_{enhance}^{(i)} \leftarrow Q_{coarse} \oplus s_i$ via Module $\Pi$} 
        \PState{$T_i \leftarrow$ Retrieve top-$k$ tools based on semantic similarity $sim(q_{enhance}^{(i)}, \mathbf{H}^{(k)})$}
        \PState{$Pool_T \leftarrow Pool_T \cup T_i$}
    \EndFor
    \PState{$Pool_T \leftarrow \text{RRF\_Ranking}(Pool_T)$} 
    \PState{$Result, P \leftarrow \mathcal{A}(S, Pool_T)$ \hfill \textit{/* Execute via task execution module */}}
    
    \Statex \qquad \textbf{// Phase 3: Online Tool Graph Evolution}
    \If{$Result$ is Success}
        \PState{Update topology of $G$ using trajectory $P$ via Module $\mathcal{U}$} 
        \PState{Re-compute pragmatic features $\mathbf{H}^{(k)}$ using the updated graph $G$}
    \EndIf
\EndFor

\PState{\Return evolved graph $G$ and features $\mathbf{H}^{(k)}$}
\end{algorithmic}
\end{algorithm}

\section{Experiment}
\label{sec:experiments}
This section delineates the experimental setup and evaluation metrics employed to assess the proposed bidirectional semantic complementary tool retrieval framework. The efficacy of the framework is rigorously validated on the specialized remote sensing (RS) dataset \textbf{GeoPlan-bench} \cite{geoplan} and the general-purpose agent benchmark \textbf{API-Bank} \cite{api-bank}. A comprehensive performance analysis is conducted by comparing our method against baselines and performing extensive ablation studies to verify the contribution of each core component.

\subsection{Experimental Setup}
\label{subsec:experimental_setup}

To evaluate the performance of the proposed method in both vertical and general-purpose domains, experiments are conducted on two distinct datasets.

\subsubsection{Datasets}
The statistical details of the datasets used in our experiments are summarized in Table \ref{tab:dataset_stats}. 

\begin{itemize}
    \item \textbf{GeoPlan-bench} \cite{geoplan}: This dataset is specifically designed for remote sensing (RS) agents, comprising diverse user queries paired with canonical tool invocation chains. It contains 248 training trajectories and 996 test trajectories, with an average tool-chain length of 10.67. Based on the sequence length and logical dependency complexity, the test set is partitioned into three difficulty levels: \textit{Simple}, \textit{Medium}, and \textit{Complex}. This tiered structure is intended to comprehensively evaluate the robustness of retrieval agents in handling professional RS tasks. Notably, due to the inherent complexity of verifying diverse geospatial products, GeoPlan-bench is designed exclusively for process-level evaluation via tool-chain fidelity, lacking a mechanism for objective end-result assessment.
    \begin{table}[htbp]
\centering
\caption{Statistical Details of the Datasets}
\label{tab:dataset_stats}
\begin{tabular}{lccccc}
\toprule
\textbf{Dataset} & \begin{tabular}[c]{@{}c@{}}\textbf{Train}\\\textbf{Size}\end{tabular} & \begin{tabular}[c]{@{}c@{}}\textbf{Test}\\\textbf{Size}\end{tabular} & \begin{tabular}[c]{@{}c@{}}\textbf{Tool}\\\textbf{Size}\end{tabular} & \begin{tabular}[c]{@{}c@{}}\textbf{Max}\\\textbf{Length}\end{tabular} & \begin{tabular}[c]{@{}c@{}}\textbf{Avg.}\\\textbf{Length}\end{tabular} \\ \midrule
GeoPlan-bench    & 248                                                                   & 996                                                                  & 104                                                                   & 20                                                                    & 10.67                                                                 \\
API-Bank\textsubscript{Level-3}         & --                                                                    & 50                                                                   & 70                                                                    & 5                                                                     & 2.62                                                                \\ \bottomrule
\end{tabular}
\end{table}
    \item \textbf{API-Bank} \cite{api-bank}: To verify the extensibility of our method from the highly coupled RS vertical domain to general-purpose scenarios, we incorporate the API-Bank benchmark. Since the APIs in the training and test sets of API-Bank are disjoint, we implement a cold-start initialization (no prior training trajectories) for the tool graph during testing. We specifically focus on Level-3 tasks, which involve long-range reasoning and require agents to perform sophisticated logical planning across multi-step invocations. The Level-3 test set includes a library of 70 general-purpose APIs. Notably, to ensure a fair evaluation of our proposed retrieval mechanism, we excluded the original retrieval tools provided in the benchmark.
\end{itemize}

\subsubsection{Baselines}
To demonstrate the superiority of the proposed framework, we compare it against traditional retrieval paradigms and graph-based methods:
\begin{itemize}
    \item \textbf{BM25}\cite{bm25}: A sparse retrieval algorithm based on Term Frequency-Inverse Document Frequency (TF-IDF), representing traditional keyword-based matching.
    \item \textbf{Dense}\cite{dense}: A mainstream semantic retrieval approach that performs cosine similarity matching using dense vectors generated by a pre-trained embedding model (\texttt{nomic-embed-text:latest}).
    \item \textbf{Graph RAG-Tool Fusion} \cite{graph-toolfusion}: Notably, several SOTA works in graph-based tool retrieval \cite{toolnet,tgr,graph-toolfusion} have not yet released their official implementations. For a persuasive comparison, we selected the representative Graph RAG-Tool Fusion and strictly followed the architectural details in its original paper for a refined reproduction. The reproduced component was integrated into our agent architecture for evaluation.
    \item \textbf{Query Rewriting}\cite{toolreagt}: A baseline that utilizes LLM-based query reformulation to align user intent with tool descriptions. While this method mitigates basic semantic mismatch, it lacks the structural task-decomposition capability required for multi-step RS workflows.
\end{itemize}
\begin{table*}[t]
\centering
\caption{Performance Comparison on GeoPlan-bench Across Different Difficulty Levels. The best results are highlighted in \textbf{bold}.}
\label{tab:geoplan_results}
\renewcommand{\arraystretch}{1.2}
\begin{tabular}{ll ccccc}
\toprule
\textbf{Difficulty} & \textbf{Method} & \textbf{Recall@k} & \textbf{NDCG@k} & \textbf{F1-Score} & \textbf{Call Recall} & \textbf{Similarity (LCS)} \\
\midrule
\multirow{5}{*}{Simple} & BM25 & 0.1202 & 0.1400 & 0.1268 & 0.1081 & 0.0934 \\
 & Dense & 0.2114 & 0.2512 & 0.3178 & 0.1889 & 0.1433 \\
 & Graph-based & 0.5184 & 0.5334 & 0.5259 & 0.2655 & 0.1779  \\
 & Query Rewriting & 0.7799 & 0.6755 & 0.7277 & 0.6511 & 0.3262 \\
 & \textbf{Ours} & \textbf{0.8191} & \textbf{0.7093} & \textbf{0.7548} & \textbf{0.7235} & \textbf{0.3511} \\
\midrule
\multirow{5}{*}{Medium} & BM25 & 0.1305 & 0.1593 & 0.1410 & 0.1145 & 0.0921 \\
 & Dense & 0.1816 & 0.2859 & 0.2338 & 0.1516 & 0.1128 \\
 & Graph-based & 0.5150 & 0.5434 & 0.5225 & 0.2469 & 0.1592 \\
 & Query Rewriting & 0.7269 & 0.6296 & 0.6702 & 0.6213 & 0.2903 \\
 & \textbf{Ours} & \textbf{0.7295} & \textbf{0.6519} & \textbf{0.6832} & \textbf{0.6304} & \textbf{0.2994} \\
\midrule
\multirow{5}{*}{Complex} & BM25 & 0.1189 & 0.1579 & 0.1342 & 0.0945 & 0.0825 \\
 & Dense & 0.1830 & 0.3073 & 0.2452 & 0.1523 & 0.1119 \\
 & Graph-based & 0.4828 & 0.5408 & 0.5054 & 0.2014 & 0.1496 \\
 & Query Rewriting & 0.6713 & 0.6223 & 0.6435 & 0.5438 & 0.2948 \\
 & \textbf{Ours} & \textbf{0.6970} & \textbf{0.6666} & \textbf{0.6790} & \textbf{0.5826} & \textbf{0.3205} \\
\midrule
\multirow{5}{*}{\textbf{Overall}} & BM25 & 0.1232 & 0.1524 & 0.1340 & 0.1057 & 0.0893 \\
 & Dense & 0.1920 & 0.2815 & 0.2656 & 0.1643 & 0.1227 \\
 & Graph-based & 0.5054 & 0.5392 & 0.5179 & 0.2379 & 0.1622 \\
 & Query Rewriting & 0.7260 & 0.6425 & 0.6805 & 0.6054 & 0.3038 \\
 & \textbf{Ours} & \textbf{0.7485} & \textbf{0.6759} & \textbf{0.7057} & \textbf{0.6455} & \textbf{0.3237} \\
\bottomrule
\end{tabular}
\end{table*}

\subsection{Evaluation Metrics}
\label{subsec:metrics}
To comprehensively evaluate the proposed framework, we adopt metrics from two complementary perspectives: \textbf{retrieval performance} and \textbf{tool invocation effectiveness}. We specifically focus on process-oriented metrics rather than end-to-end outputs for three key reasons. First, evaluating the fidelity of the \textbf{tool-calling trajectory} directly validates the core contribution of our BSCTR framework in addressing semantic misalignment and workflow planning. Second, in the remote sensing (RS) domain, the \textbf{logical integrity of the tool chain} is paramount; strict causal dependencies (e.g., atmospheric correction preceding feature extraction) mean that workflow errors inevitably dictate flawed outputs. Finally, since \textbf{GeoPlan-Bench} is designed for process-level validation, this focus allows us to objectively isolate the agent's reasoning capabilities from the performance variances of underlying image-processing algorithms.
\subsubsection{Retrieval Performance Metrics}
\begin{itemize}
    \item \textbf{Recall@$k$}: Measures the proportion of ground-truth tools successfully included within the top-$k$ recalled subset.
    \item \textbf{NDCG@$k$}: Normalized Discounted Cumulative Gain at rank $k$. It evaluates the ranking quality of the retrieval results, focusing on whether the correct tools are prioritized at higher positions in the list.
    \item \textbf{F1 Score}: The harmonic mean of retrieval precision and recall, providing a balanced assessment of the retrieval accuracy.
\end{itemize}

\subsubsection{Tool Invocation Performance Metrics}
\begin{itemize}
    \item \textbf{Tool Call Recall}: Defined as the ratio of correctly invoked tools in the agent's actual execution trajectory to the total number of tools in the ground-truth tool-chain.
    \item \textbf{Tool Trajectory Similarity}: This metric reflects the agent's ability to understand the logical dependencies between tools. It is calculated as the normalized score of the Longest Common Subsequence (LCS) between the predicted trajectory $P$ and the ground-truth trajectory $T$:
    \begin{equation}
    \text{Sim}(P, T) = \frac{|\text{LCS}(P, T)|}{\max(|P|, |T|)}
    \label{eq:lcs}
    \end{equation}
    where $|\text{LCS}(P, T)|$ denotes the length of the longest common subsequence. This score quantifies the structural alignment of the execution sequence, indicating whether the agent correctly follows the domain-specific workflow.
\end{itemize}

\subsection{Implementation Details}
In the proposed BSCTR framework, the \textit{Predecessor mode} is adopted for tool graph aggregation, with the smoothing factor $\alpha$ set to $0.5$ and the aggregation order $k$ set to $1$. 
All modules requiring LLM reasoning are powered by the \textbf{DeepSeek-V3.2} model \cite{deepseek}, and the \texttt{nomic-embed-text:latest} model is unified as the embedding backbone to maintain consistent semantic space alignment.

For the evaluation metrics (Recall@$k$ and NDCG@$k$), we employ two complementary settings. 
First, we set $k$ to the \textbf{maximum trajectory length} ($k=20$ for GeoPlan-bench and $k=5$ for API-Bank\textsubscript{Level-3}). 
This is justified by the \textit{minimal sufficiency} principle: in specialized RS workflows, the agent must have access to all requisite tools in a sequence to maintain the integrity of professional logic. 
Furthermore, since the dynamic $k$ values used in the original GeoPlan-bench \cite{geoplan} were not disclosed, using the maximum length ensures \textit{reproducibility}. 

Second, specifically for the \textbf{GeoPlan-bench} dataset, we introduce an additional set of experiments with a broader spectrum of retrieval scales, where $k \in \{5, 10, 15\}$. 
The objective of this multi-$k$ analysis is to enhance the rigor of our evaluation by investigating the framework's sensitivity to varying search space constraints. 
By exploring these intermediate $k$ values, we can more precisely elucidate the framework's ability to prioritize relevant tools at the top of the candidate list, thereby validating its ranking precision and robustness across different operational requirements.

\subsection{GeoPlan-bench Results}
\label{subsec:geoplan_results}

\subsubsection{Performance Across Difficulty Levels}
\label{subsubsec:difficulty_results}
Table \ref{tab:geoplan_results} summarizes the performance comparison of various retrieval methods on the GeoPlan-bench dataset. Overall, the proposed \textbf{BSCTR} framework consistently outperforms all baselines across all difficulty levels. Specifically, it achieves an overall \textit{Recall@k} of \textbf{0.7485} and a \textit{Similarity (LCS)} of \textbf{0.3237}, marking a significant improvement over traditional paradigms.

The superior performance is primarily attributed to BSCTR's ability to bridge the dual-sided semantic misalignment. On the \textbf{query side}, RS intentions are often macro-level and abstract (e.g., ``analyze crop health''), lacking the specific \textit{functional semantics} required to trigger specialized tools. Traditional methods like BM25 and Dense retrieval fail significantly in these scenarios (Recall@$k < 0.22$), as they rely on literal keyword matching rather than intent deconstruction. Our planning-based enhancement translates these intentions into explicit technical sub-tasks, providing the necessary ``logical hooks.'' 

On the \textbf{tool side}, while standard documentation provides technical details, it lacks \textit{contextual semantics}---the pragmatic dependencies within a specialized workflow. By comparing our method with the \textit{Query Rewriting} baseline, we observe that even when query functionality is supplemented, the lack of tool-side context leads to inferior execution (lower Call Recall). BSCTR ensures that the agent retrieves not just functionally relevant tools, but contextually appropriate ones, resulting in a \textit{Call Recall} of \textbf{0.6455}.

\begin{table}[htbp]
\centering
\caption{Overall Performance Comparison on GeoPlan-bench Across Different Retrieval Scales $k$. The best results are highlighted in \textbf{bold}.}
\label{tab:k_comparison}
\setlength{\tabcolsep}{3.5pt}
\small 
\begin{tabular}{l cc cc cc}
\toprule
\multirow{2}{*}{\textbf{Method}} & \multicolumn{2}{c}{\textbf{k = 5}} & \multicolumn{2}{c}{\textbf{k = 10}} & \multicolumn{2}{c}{\textbf{k = 15}} \\ 
\cmidrule(lr){2-3} \cmidrule(lr){4-5} \cmidrule(lr){6-7}
 & \textbf{R@k} & \textbf{N@k} & \textbf{R@k} & \textbf{N@k} & \textbf{R@k} & \textbf{N@k} \\ \midrule
BM25             & 0.1236 & 0.2458 & 0.1236 & 0.1668 & 0.1236 & 0.1512 \\
Dense            & 0.1954 & 0.4838 & 0.1954 & 0.3326 & 0.1954 & 0.3058 \\
Graph-based      & 0.2700 & 0.5990 & 0.4411 & 0.5443 & 0.4833 & 0.5261 \\
Query Rewriting  & 0.2629 & 0.5448 & 0.4660 & 0.5312 & 0.6484 & 0.5971 \\ \midrule
\textbf{Ours}    & \textbf{0.2897} & \textbf{0.6053} & \textbf{0.5052} & \textbf{0.5821} & \textbf{0.6752} & \textbf{0.6363} \\ \bottomrule
\end{tabular}
\end{table}

\subsubsection{Impact of Retrieval Scale $k$}
\label{subsubsec:k_analysis}
To evaluate the robustness and ranking precision of the framework, we conducted a multi-scale retrieval analysis on GeoPlan-bench, as shown in Table \ref{tab:k_comparison}. At a highly constrained scale of \textbf{$k=5$}, our method achieves an NDCG of \textbf{0.6053}, significantly surpassing the Query Rewriting (0.5448) and Graph-based (0.5990) baselines. This high ranking precision at small $k$ values demonstrates that our bidirectional mechanism effectively prioritizes the most relevant tools at the very top of the candidate list, which is vital for RS agents operating within limited context windows to avoid information decay.

Moreover, as $k$ increases from 5 to 15, BSCTR exhibits steady and substantial growth, with Recall increasing from \textbf{0.2897} to \textbf{0.6752}. This trend underscores that our approach can uncover deep-seated tool dependencies and sequential constraints that are completely invisible to literal or shallow semantic matching.

\begin{table*}[t]
\centering
\caption{Performance Comparison on the General-Purpose API-Bank Dataset (Level-3). The best results are highlighted in \textbf{bold}.}
\label{tab:apibank_results}
\renewcommand{\arraystretch}{1.2}
\begin{tabular}{l ccccc}
\toprule
\textbf{Method} & \textbf{Recall@k} & \textbf{NDCG@k} & \textbf{F1-Score} & \textbf{Call Recall} & \textbf{Similarity (LCS)} \\
\midrule
BM25 & 0.5900 & 0.6643 & 0.6272 & 0.4167 & 0.3770 \\
Dense & 0.8567 & 0.8410 & 0.8489 & 0.6567 & 0.5710 \\
Graph-based & 0.8967 & 0.8571 & 0.8769 & 0.7767 & 0.5683 \\
Query Rewriting & 0.8981 & 0.8562 & 0.8771 & 0.7767 & 0.5940 \\
\textbf{Ours} & \textbf{0.9067} & \textbf{0.8613} & \textbf{0.8804} & \textbf{0.7900} & \textbf{0.6090} \\
\bottomrule
\end{tabular}
\end{table*}

\subsubsection{Discussion on Execution Metrics}
\label{subsubsec:lcs_discussion}
Finally, while the absolute values of the \textit{Tool Trajectory Similarity} (LCS) might appear relatively modest (peaking at 0.3237), this is a well-known phenomenon in long-horizon agent evaluations and can be attributed to three primary factors: 
\begin{itemize}
    \item \textbf{Exploratory nature of the ReAct paradigm:} ReAct agents \cite{react} inherently engage in trial-and-error, environmental observation, and self-correction. These auxiliary reasoning steps inflate the predicted trajectory length, which heavily penalizes the normalized LCS formula (where $\max(|P|, |T|)$ serves as the denominator).
    \item \textbf{Workflow equifinality:} RS analytical tasks often permit multiple valid operational sequences (e.g., swapping certain commutative preprocessing steps) to achieve the same scientific goal. LCS strictly evaluates matching against a single canonical ground-truth path, thus yielding lower absolute scores for valid alternative solutions.
    \item \textbf{Compounding divergence in long chains:} With an average ground-truth length exceeding 10 steps in GeoPlan-bench, a single divergent step early in the sequence can drastically truncate the common subsequence. 
\end{itemize}
Despite these challenges, BSCTR achieves the highest LCS across all categories, proving its superior ability to capture the sequential dependencies essential for professional RS workflows.

\subsection{API-Bank Results}
\label{subsec:apibank_analysis}

To further validate the cross-domain generalization and extensibility of the proposed framework, we conducted evaluations on the general-purpose benchmark \textbf{API-Bank}\textsubscript{Level-3}. As summarized in Table \ref{tab:apibank_results}, our \textbf{Bidirectional Semantic Complementary Tool Retrieval} method consistently achieves the state-of-the-art (SOTA) performance across all evaluation metrics, with a \textit{Recall@k} of \textbf{0.9067} and an \textit{F1-Score} of \textbf{0.8804}.

The superior performance on API-Bank demonstrates the universal applicability of our bidirectional semantic complementary mechanism. Although the tool chains in general domains (e.g., calendar management, email services, and weather queries) are typically shorter than those in remote sensing (RS), they still suffer from the fundamental challenge of ``semantic asymmetry.'' Specifically, user queries in daily scenarios are often highly implicit, while API documentation remains technical and disjointed. Our results show that by supplementing functional semantics on the query side and injecting contextual dependencies on the tool side, the agent can achieve an \textit{NDCG@k} of \textbf{0.8613}, significantly outperforming the \textit{Query Rewriting} (0.8562) and \textit{Graph-based} (0.8571) baselines. This indicates that the synergy between intent deconstruction and topological context is a robust solution for tool retrieval, regardless of the specific domain.

Furthermore, the high \textit{Similarity (LCS)} score of \textbf{0.6090} and \textit{Call Recall} of \textbf{0.7900} reflect the agent's exceptional ability to handle multi-step reasoning in general scenarios. Compared to the RS domain, the relatively higher LCS scores on API-Bank suggest that the tool dependencies in general tasks are more straightforward; however, our method still provides the most accurate sequence alignment. This proves that the proposed framework is not merely a niche solution for specialized RS tasks, but a versatile retrieval paradigm capable of effectively bridging the semantic gap in diverse, open-domain tool-use environments. The robust performance in a cold-start setting (without pre-training trajectories on API-Bank) further underscores the adaptability and plug-and-play nature of our dynamic tool graph evolution mechanism.

\subsection{Ablation Study} 
\label{subsec:ablation}
To systematically evaluate the individual contributions of the two core components in the Bidirectional Semantic Complementary Tool Retrieval framework---Planning-based Query Enhancement and Tool Dependency Graph Modeling---we conducted ablation experiments on both API-Bank and GeoPlan-bench. The results are presented in Table \ref{tab:ablation}.
\begin{table*}[t]
\centering
\caption{Ablation Study Results on API-Bank and GeoPlan-bench.}
\label{tab:ablation}
\renewcommand{\arraystretch}{1.2}
\begin{tabular}{l l ccccc}
\toprule
\textbf{Dataset} & \textbf{Configuration} & \textbf{Recall@k} & \textbf{NDCG@k} & \textbf{F1-Score} & \textbf{Call Recall} & \textbf{Similarity (LCS)} \\ \midrule
\multirow{3}{*}{\textbf{API-Bank}\textsubscript{Level-3}} & w/o Planning & 0.8567 & 0.8374 & 0.8445 & 0.6433 & 0.5587 \\
 & w/o Tool Graph & 0.8981 & 0.8562 & 0.8771 & 0.7767 & 0.5940 \\
 & \textbf{Full (Ours)} & \textbf{0.9067} & \textbf{0.8613} & \textbf{0.8804} & \textbf{0.7900} & \textbf{0.6090} \\ \midrule
\multirow{3}{*}{\textbf{GeoPlan-bench}} & w/o Planning & 0.1883 & 0.2968 & 0.2426 & 0.1662 & 0.1203 \\
 & w/o Tool Graph & 0.7506 & 0.6490 & 0.6913 & 0.6054 & 0.3074 \\
 & \textbf{Full (Ours)} & \textbf{0.7485} & \textbf{0.6759} & \textbf{0.7057} & \textbf{0.6455} & \textbf{0.3237} \\ \bottomrule
\end{tabular}
\end{table*}

\subsubsection{Global Synergy Analysis}
The experimental results indicate that removing either the query-side enhancement or the tool-side graph injection leads to a decline in performance across all metrics, validating the synergy of our bi-directional semantic complementary mechanism. Notably, the removal of the planning module (\textit{w/o Planning}) leads to a much more catastrophic performance collapse on GeoPlan-bench (Recall@$k$ dropping from 0.7485 to 0.1883) compared to API-Bank. This disparity underscores the unique challenges of the remote sensing (RS) domain.

\subsubsection{The Critical Role of Planning in RS Domain}
The significant performance gap observed when omitting the planning module in the RS domain, as opposed to the general domain, is driven by the following factors:
\begin{itemize}
    \item \textbf{Implicit vs. Explicit Semantics:} In general domains (e.g., API-Bank), user queries such as ``check the weather'' share high lexical and semantic overlap with API descriptions. Conversely, RS queries are typically formulated as high-level scientific objectives (e.g., ``analyze forest cover changes'') that share almost no common keywords with specialized tools like ``Atmospheric Correction'' or ``NDVI Calculation.'' Without planning-based decomposition, the retriever fails to bridge this vast semantic chasm, as the macro-intent provides insufficient functional cues for matching.
    
    \item \textbf{Technical Granularity and Expertise Translation:} RS tasks involve rigid professional workflows where essential pre-processing steps (e.g., geometric calibration, radiometric correction) are mandatory yet never explicitly mentioned in user queries. Planning acts as an \textit{expertise translator}, converting abstract intentions into a sequence of fine-grained technical sub-tasks. By uncovering these implicit functional requirements, the planning module ensures that the retriever can identify the professional tool-chain necessary to satisfy the underlying scientific logic, which is a prerequisite for successful retrieval in specialized vertical domains.
\end{itemize}

\subsubsection{Impact of Tool Graph Modeling}
The removal of the tool graph (\textit{w/o Tool Graph}) primarily impacts the \textit{Similarity (LCS)} and \textit{Call Recall} metrics. This confirms that even with an enhanced query, retrieval may still struggle to distinguish between functionally similar tools without \textbf{contextual semantics}. The tool graph injects ``workflow context'' into node representations, ensuring that retrieved tools are not only functionally relevant but also logically consistent with the preceding steps in the RS workflow. This component serves as the ``logical anchor'' that maintains the structural integrity of long-chain task execution.

\subsection{Parameter Sensitivity Analysis} 
\label{subsec:parameter_k}
In this section, we investigate the impact of two key hyperparameters in the tool graph modeling module: the aggregation order $k$ and the smoothing factor $\alpha$. All experiments in this subsection are conducted on the GeoPlan-bench dataset.

\subsubsection{Impact of Aggregation Order $k$}
The aggregation order $k$ determines the range of the tool-chain context injected into the current node representation. As summarized in Fig. \ref{fig:sensitivity_k}, we evaluate the retrieval performance (Recall) by varying $k$ from 1 to 5. 

The results indicate that the optimal performance (0.7485) is achieved at $k=1$. As $k$ increases, we observe a gradual degradation in retrieval accuracy. This phenomenon can be attributed to the \textit{over-smoothing} effect commonly found in graph neural networks. Specifically, while a higher-order $k$ allows the node to capture long-range dependencies, it also introduces noise from distant, less-relevant tools in the workflow, thereby diluting the distinct functional features of the current tool. Therefore, we set $k=1$ for our BSCTR framework to maintain a balance between local specificity and immediate contextual relevance.

\definecolor{deepblue}{RGB}{0, 102, 204}
\definecolor{lightblue}{RGB}{173, 216, 230}
\definecolor{gridgray}{RGB}{220, 220, 220}

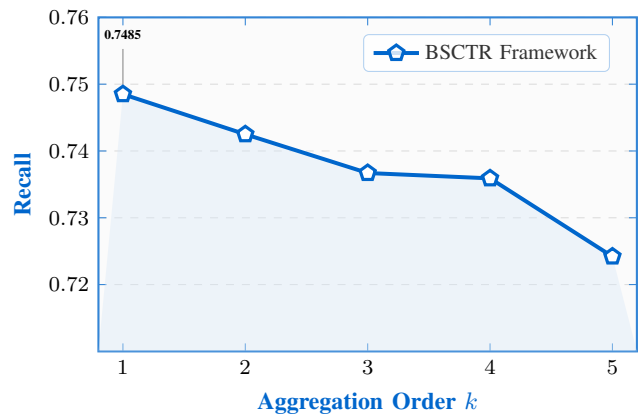
\begin{figure}[htbp]
\centering
\begin{tikzpicture}
    \begin{axis}[
        width=0.48\textwidth,
        height=6cm,
        axis background/.style={fill=gray!3},
        tick align=inside,
        major tick length=3pt,
        axis line style={deepblue!80, thick},
        tick style={deepblue!80, thin},
        xmajorgrids=false,
        ymajorgrids=true,
        grid style={dashed, gridgray},
        xlabel={\textbf{Aggregation Order} $k$},
        ylabel={\textbf{Recall}},
        xlabel style={font=\small\color{deepblue}},
        ylabel style={font=\small\color{deepblue}},
        tick label style={font=\footnotesize},
        xmin=0.8, xmax=5.2,
        ymin=0.71, ymax=0.76,
        xtick={1,2,3,4,5},
        ytick={0.72, 0.73, 0.74, 0.75, 0.76},
        yticklabel style={/pgf/number format/fixed, /pgf/number format/precision=3},
        legend style={
            at={(0.95,0.95)},
            anchor=north east,
            nodes={scale=0.8, transform shape},
            draw=deepblue!30,
            fill=white,
            fill opacity=0.8,
            rounded corners=2pt
        }
    ]

    \addplot[name path=line, deepblue, thick, mark=none, forget plot] 
    coordinates {(1, 0.7485) (2, 0.7425) (3, 0.7367) (4, 0.7359) (5, 0.7242)};
    
    \path[name path=bottom] (axis cs:0.8,0.71) -- (axis cs:5.2,0.71);
    
    \addplot[deepblue!15, opacity=0.4, forget plot] fill between[of=line and bottom];

    \addplot[
        color=deepblue,
        mark=pentagon*, 
        mark size=3pt,
        line width=1.5pt,
        mark options={fill=white, draw=deepblue, line width=1.2pt} 
    ]
    coordinates {
        (1, 0.7485)
        (2, 0.7425)
        (3, 0.7367)
        (4, 0.7359)
        (5, 0.7242)
    };
    \addlegendentry{BSCTR Framework}

    \node[pin={90:{\tiny \textbf{0.7485}}}, color=deepblue!80] at (axis cs:1, 0.7485) {};

    \end{axis}
\end{tikzpicture}
\caption{Impact of the aggregation order $k$ on retrieval performance. The shaded area represents the performance confidence trend, with the peak occurring at $k=1$, highlighting the importance of local contextual features.}
\label{fig:sensitivity_k}
\end{figure}

\subsubsection{Impact of Smoothing Factor $\alpha$}
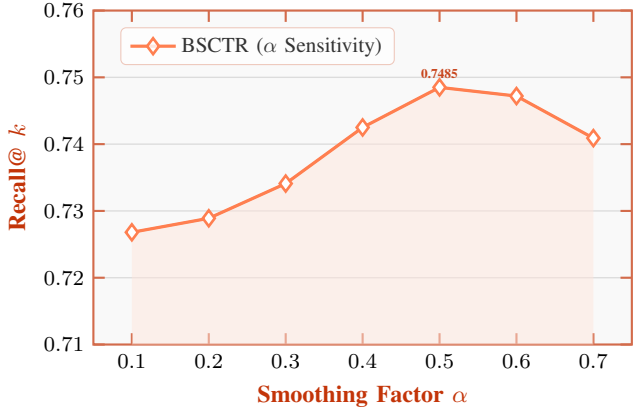
\begin{figure}[htbp]
\centering
\begin{tikzpicture}
    \begin{axis}[
        width=0.48\textwidth,
        height=6cm,
        axis background/.style={fill=gray!5},
        axis line style={darkorange!80, thick},
        tick style={darkorange!80, thick},
        tick align=inside,
        xmajorgrids=false,
        ymajorgrids=true,
        grid style={color=gridgray, line width=0.5pt},
        xlabel={\textbf{Smoothing Factor} $\alpha$},
        ylabel={\textbf{Recall@} $k$},
        xlabel style={font=\small\color{darkorange!110}},
        ylabel style={font=\small\color{darkorange!110}},
        tick label style={font=\footnotesize},
        xmin=0.05, xmax=0.75,
        ymin=0.71, ymax=0.76,
        xtick={0.1, 0.2, 0.3, 0.4, 0.5, 0.6, 0.7},
        ytick={0.71, 0.72, 0.73, 0.74, 0.75, 0.76},
        scaled y ticks=false,
        yticklabel style={/pgf/number format/fixed, /pgf/number format/precision=3},
        legend style={
            at={(0.05,0.95)},
            anchor=north west,
            nodes={scale=0.8, transform shape},
            draw=darkorange!30,
            fill=white,
            fill opacity=0.8,
            rounded corners=2pt
        }
    ]

    \addplot[name path=top, myorange, thick, mark=none, forget plot] 
    coordinates {
        (0.1, 0.7268)
        (0.2, 0.7289)
        (0.3, 0.7341)
        (0.4, 0.7425)
        (0.5, 0.7485)
        (0.6, 0.7472)
        (0.7, 0.7409)
    };
    
    \path[name path=bottom] (axis cs:0.1,0.71) -- (axis cs:0.7,0.71);
    
    \addplot[myorange!20, opacity=0.5, forget plot] fill between[of=top and bottom];

    \addplot[
        color=myorange,
        mark=diamond*, 
        mark size=3pt,
        line width=1.2pt,
        mark options={fill=white, draw=myorange, line width=1pt}
    ]
    coordinates {
        (0.1, 0.7268)
        (0.2, 0.7289)
        (0.3, 0.7341)
        (0.4, 0.7425)
        (0.5, 0.7485)
        (0.6, 0.7472)
        (0.7, 0.7409)
    };
    \addlegendentry{BSCTR ($\alpha$ Sensitivity)}

    \node[above, font=\tiny\bfseries, color=darkorange] at (axis cs:0.5, 0.7485) {0.7485};

    \end{axis}
\end{tikzpicture}
\caption{Parameter sensitivity analysis of the smoothing factor $\alpha$ on the GeoPlan-bench dataset.}
\label{fig:sensitivity_alpha}
\end{figure}

The smoothing factor $\alpha$ is a critical parameter that balances the preservation of a tool's \textit{functional semantics} (from its original documentation) and the injection of \textit{contextual semantics} (from its topological neighborhood). As illustrated in Fig. \ref{fig:sensitivity_alpha}, we evaluate the retrieval performance by varying $\alpha$ from 0.1 to 0.7.

The experimental results exhibit a clear bell-shaped trend, with the Recall@$k$ peaking at \textbf{0.7485} when $\alpha = 0.5$. This phenomenon can be analyzed from two perspectives:
\begin{itemize}
    \item \textbf{Insufficient Context Aggregation ($\alpha < 0.5$):} At lower values of $\alpha$, the tool representation is dominated by its original textual description. Due to the inherent ``semantic asymmetry'' between abstract user queries and technical documentation, the individual functional features are insufficient to provide enough discriminative cues for complex remote sensing (RS) workflows. The lower performance in this range suggests that the information gain from the tool dependency graph is inadequate to bridge the semantic gap.
    \item \textbf{Feature Homogenization and Over-smoothing ($\alpha > 0.5$):} As $\alpha$ exceeds the optimal threshold, a significant performance degradation is observed. This is attributed to the \textit{over-smoothing} effect, where excessive topological information begins to overshadow the tool's unique functional identity. In the high-dimensional embedding space, an over-reliance on neighborhood aggregation leads to \textit{feature homogenization}, causing semantically distinct tools within the same tool-chain to become indistinguishable. Consequently, the retriever loses its ability to pinpoint the exact tool required for a specific sub-task, leading to a drop in Recall.
\end{itemize}

In summary, $\alpha = 0.5$ represents the optimal equilibrium, where the tool retains its specific functional definition while being sufficiently enriched by the pragmatic context of the RS tool-chain.

\subsection{Analysis of Graph Propagation Directions} 
\label{subsec:propagation_direction}
\begin{figure}[htbp]
\centering
\begin{tikzpicture}
    \begin{axis}[
        ybar, 
        enlarge x limits=0.25,
        width=0.48\textwidth,
        height=6cm,
        axis background/.style={fill=gray!3},
        tick align=inside,
        ylabel={Recall@$k$},
        xlabel={Propagation Mode ($\omega$)},
        symbolic x coords={p, s, sym}, 
        xtick=data,
        nodes near coords, 
        nodes near coords style={
            font=\tiny, 
            /pgf/number format/fixed, 
            /pgf/number format/precision=4,
            /pgf/number format/zerofill 
        },
        every node near coord/.append style={yshift=2pt},
        ymin=0.72, ymax=0.76, 
        bar width=22pt,
        ymajorgrids=true,
        grid style={dashed, gray!30},
        axis line style={tealblue!80, thick},
        tick label style={font=\small},
        xticklabels={Predecessor, Successor, Symmetric},
        x tick label style={font=\footnotesize},
    ]

    \addplot[
        draw=tealblue,
        fill=tealblue!60,
        postaction={
            pattern=north east lines, 
            pattern color=tealblue!20
        }
    ] 
    coordinates {(p, 0.7486) (s, 0.7362) (sym, 0.7485)};
    
    \end{axis}
\end{tikzpicture}
\caption{Comparison of different graph propagation directions on GeoPlan-bench. Mode $p$ (Predecessor) and Mode $sym$ (Symmetric) exhibit superior performance compared to Mode $s$ (Successor), underscoring the vital role of historical context in RS tool-chains.}
\label{fig:direction_analysis}
\end{figure}
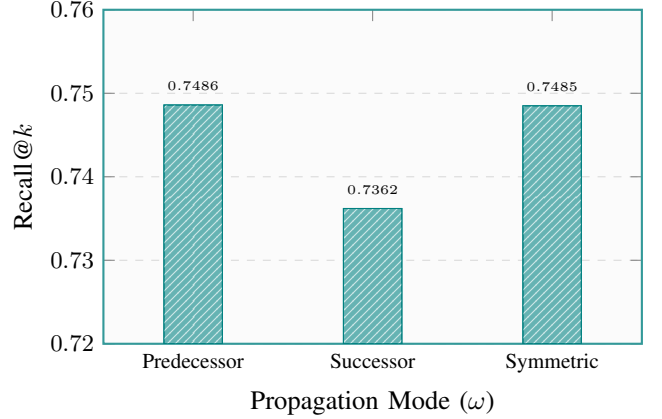

In this subsection, we investigate the impact of different graph propagation directions ($\omega$), as defined in Section \ref{subsec:tool_graph_modeling}, to determine the optimal orientation for injecting contextual semantics. The experimental results, illustrated in Fig. \ref{fig:direction_analysis}, reveal that the \textbf{Predecessor mode} ($\omega = p$, 0.7486) and \textbf{Symmetric mode} ($\omega = sym$, 0.7485) significantly outperform the \textbf{Successor mode} ($\omega = s$, 0.7362).

This performance disparity provides critical insights into the logic of remote sensing (RS) workflows:
\begin{itemize}
    \item \textbf{Primacy of Historical Context:} The superiority of Mode $p$ suggests that in RS tool-chains, the functional identity of a tool is predominantly defined by its ``upstream'' prerequisites. For example, the execution of feature extraction is logically dependent on prior radiometric and geometric corrections. By explicitly aggregating precursor information, the current tool node captures the necessary \textit{causal constraints}, making it more distinguishable during retrieval.
    \item \textbf{Inconsistency of Future Context:} The noticeably lower performance of Mode $s$ indicates that aggregating information from subsequent (downstream) tools may introduce ``topological noise.'' Since a single precursor tool might lead to multiple divergent downstream paths, future context provides less restrictive semantic cues for the current retrieval step compared to the deterministic history of the workflow.
    \item \textbf{Symmetric Redundancy:} While the Symmetric mode ($sym$) achieves nearly identical performance to the Predecessor mode ($p$), it does not provide marginal gains. This implies that historical context alone is sufficient to bridge the semantic gap, and the inclusion of successor information offers redundant cues in the context of RS task planning.
\end{itemize}

Consequently, we adopt the \textbf{Predecessor mode} as the default configuration for the BSCTR framework, as it captures the core causal logic of professional RS workflows with high precision.

\section{Discussion}
\subsection{The Necessity of Bidirectional Complementarity}
The experimental evidence from both GeoPlan-bench and API-Bank underscores that the ``semantic asymmetry'' in tool retrieval cannot be fully resolved by optimizing only one side of the retrieval equation. While query enhancement provides the necessary \textit{functional hooks}, it is the tool dependency graph that offers the \textit{contextual glue} required to hold a long-chain workflow together. In the remote sensing domain, where tasks involve high levels of technical abstraction, our bidirectional approach creates a shared semantic space that successfully bridges the gap between non-expert natural language and expert-level technical tools.
\subsection{Causal Logic and Domain Expertise}
A key finding in our analysis of propagation directions is the superiority of the Predecessor mode ($p$). This aligns perfectly with the inherent causal logic of RS image processing, where each subsequent step is strictly constrained by the outputs of precursor operations. Our framework effectively transforms these implicit professional workflows into explicit topological features. Unlike general-purpose agents, RS agents must respect the physical and radiometric consistency of the data; by injecting historical context into tool representations, we enable the agent to ``perceive'' the underlying scientific workflow during the retrieval phase, even before the actual execution begins.
\subsection{Limitations}
Despite its superior performance, the proposed Bidirectional Semantic Complementary Tool Retrieval framework has several limitations that warrant further investigation. First, the efficacy of query-side functional enhancement is closely coupled with the LLM's initial planning accuracy; any logical hallucinations during task decomposition may propagate through the retrieval process, leading to cascading errors. Second, the dynamic evolution of the tool graph relies on high-quality execution trajectories, which may trigger cold-start challenges in data-sparse or emerging remote sensing sub-fields. Third, the current text-centric paradigm lacks direct integration of multi-modal cues---such as sensor specifications or visual metadata---which are indispensable for more nuanced, data-driven tool selection. Finally, as the tool library scales to massive proportions, the computational overhead of real-time neighborhood feature aggregation remains a potential bottleneck for high-concurrency applications.
\section{Conclusion}
This paper addresses the \textit{semantic asymmetry} in complex RS tool retrieval by proposing the \textbf{Bidirectional Semantic Complementary Tool Retrieval} framework. By synergizing planning-based query enhancement with dynamic tool dependency graphs, we effectively bridge the gap between abstract user intentions and technical documentation. 
Experimental results on \textit{GeoPlan-bench} and \textit{API-Bank} demonstrate that our framework achieves new state-of-the-art (SOTA) performance, significantly enhancing both retrieval precision and the logical integrity of execution. 

Future research will focus on integrating multi-modal visual cues and investigating the fundamental essence of graph connectivity. Furthermore, we plan to optimize information aggregation via parametric learning to further enhance tool mastery and framework efficiency in intricate geospatial environments.

\bibliographystyle{IEEEtran}
\bibliography{ref}
\end{document}